\documentstyle[aps,multicol,amssymb,psfig,float,times]{revtex}

\newcommand{\be}{\begin{equation}}
\newcommand{\ee}{\end{equation}}
\newcommand{\bea}{\begin{eqnarray}}
\newcommand{\eea}{\end{eqnarray}}

\def\p1{\pi_1}

\def\l{\lambda}
\def\f{\phi}

\def\r{\rho}

\def\ep{\epsilon}

\def\pmas{\partial_+}
\def\pmen{\partial_-}

\begin{document}

\draft

\title{Evaporation of near-extremal Reissner-Nordstr\"om black holes}

\author{Alessandro Fabbri$^{\ast}$}
\address{Dipartimento di Fisica dell'Universit\`a di Bologna and INFN
sezione di Bologna,\\
Via Irnerio 46, 40126 Bologna, Italy.}
\author{Diego J. Navarro$^{\dagger}$ and Jos\'e Navarro-Salas$^{\ddagger}$}
\address{Departamento de F\'{\i}sica Te\'orica and IFIC, Centro Mixto
Universidad de Valencia-CSIC.\\
Facultad de F\'{\i}sica, Universidad de Valencia, Burjassot-46100, Valencia,
Spain.}

\maketitle

\begin{abstract}

The formation of near-extremal Reissner-Nordstr\"om black holes in the S-wave
approximation can be described, near the event horizon, by an effective
solvable model. The corresponding one-loop quantum theory remains solvable and
allows to follow analytically the evaporation process which is shown to
require an infinite amount of time.
\end{abstract}


\begin{multicols}{2}

\narrowtext

Black holes are the most fascinating objects in General Relativity. Since
Hawking discovered that they emit thermal radiation \cite{h}, it has been a
long standing puzzle to explain their thermodynamical properties, in terms of
some microscopic structure, and to understand the dynamical evolution beyond
Hawking initial scheme where the gravitational field was treated as a fixed
background. Extremal and near-extremal charged black holes have recently
played a fundamental role in String Theory, where in some special cases it has
been possible to give a statistical explanation of the Bekenstein-Hawking area
law for their entropy  \cite{svms}. Moreover, the scattering of low-energy
particles off extremal black holes also provides a convenient setting to study
the evaporation process including back-reaction effects. By throwing a
long-wavelength particle into an extremal hole, a non-extremal configuration is
created and quantum-mechanically one expects it to decay via Hawking emission
back to extremality. To render this problem tractable one can boil it down
considering large $q$ and incoming neutral matter with zero angular momentum.
In this context, dilatonic black holes \cite{ghs} were extensively considered
since the scattering particle-hole and the ensuing information loss problem
can be analyzed in an analytical framework. This is so because the problem can
be reduced to study a two-dimensional effective theory \cite{cghs} which turns
out to be solvable both at the classical and  at the one-loop quantum level
\cite{rst,h2,bpp,mik} (see also the reviews \cite{tgh} for a more detailed
description and as useful references for the methods used in this work). Among
the `nice' properties of these dilatonic black holes, making the problem under
study rather special, the extremal black holes are classically completely
regular and, moreover, the temperature near extremality
is constant. In contrast, the Reissner-Nordstr\"om (RN) black holes always
have singularities (extremal case $lm=q$ included) and the temperature goes as
$T_H \sim \sqrt{lm-q}/q^2$ near extremality. This case was reconsidered after
the improvements in the physical understanding obtained with dilatonic black
holes. However, only partial analytic (obtained by means of the adiabatic
approximation) and numerical answers have been obtained \cite{st,lo}. The
purpose of this letter is to present the first exact results on the
evaporation process of a near-extremal RN black hole. If we consider the
physics in the S-wave approximation, near the horizon we can describe it by an
effective model (the Jackiw-Teitelboim model \cite{jt}) which is solvable even
when back-reaction effects are included.\\

Let us consider the solutions of Einstein-Maxwell gravity with null infalling
matter
\bea
\label{rn2}
d\bar{s}^2 &=& -\frac{(r-r_+)(r-r_-)}{r^2} \; dv^2 + 2 dr dv +
r^2 d\Omega^2 \, , \\
F &=& q \ep_2 \, ,
\eea
where $\ep_2$ is the volume element of the unit S$^2$. The only nonzero
component of the stress tensor is given by
\be
T_{vv}=\frac{\partial_v(r_+(v)+r_-(v))}{8\pi l^2 r^2} \, ,
\ee
where $l^2=G$ is Newton's gravitational constant. One can describe the
formation of a non-extremal black hole by sending a low-energy shock wave
\be
\label{shock}
T_{vv} = \frac{\Delta m}{4\pi r^2} \delta(v-v_0) \, ,
\ee
in the extremal geometry ($v<v_0$). This model can be described by an
effective two-dimensional theory given by
\be
\label{effective}
I = \int d^2x \sqrt{-g} \left[ R\f + l^{-2} V(\f) - \frac{1}{2}
|\nabla f|^2 \right] \, ,
\ee
where the field $f$ represents the null matter with $(\partial_v f)^2\equiv
T^f_{vv}= 4\pi r^2 T_{vv}$ and
\be
\f = \frac{r^2}{4l^2} \, , \qquad
V(\f) = (4\f)^{-\frac{1}{2}} - q^2(4\phi)^{-\frac{3}{2}} \, .
\ee
The two-dimensional metric is related to the $r-v$ projection of (\ref{rn2})
by the conformal rescaling
\be
\label{conformal}
ds^2 = \sqrt{\f}d\bar{s}^2 \, .
\ee
The extremal black hole is recovered for the zero of the potential $V(\f_0)=0$,
which corresponds to $\f_0=q^2/4$. This fact suggests us to consider the
effective near-horizon and near-extremal theory defined by the expansion of
(\ref{effective}) around $\f_0$
\be
\label{e}
\f = \f_0 + \tilde{\f} \, , \qquad
m = lq + \Delta m \, .
\ee
We obtain
\be
\label{jtaction}
I = \int d^2x \sqrt{-g} \left[ (R + \frac{4}{l^2q^3}) \tilde{\f} - \frac{1}{2}
|\nabla f|^2 \right] + {\cal{O}}(\tilde{\f}^2) \, ,
\ee
where the leading order term is just the Jackiw-Teitelboim (JT) model, which
now arises as the effective theory governing the dynamics near extremality and
close to the horizon. A $d=1$ realization of the AdS$_{d+1}/$CFT$_d$
correspondence \cite{mw} in the JT model exactly accounts for the deviation
from extremality of the Bekenstein-Hawking entropy of RN black holes
\cite{nnn}. Therefore one can also expect to obtain an exact picture of the
evaporation process near the horizon. The formation of a near-extremal black
hole due to a shock wave (\ref{shock}) can be pushed down to the JT metrics of
constant negative curvature. For $v<v_0$ we have the extremal RN configuration
and its near-horizon geometry is given by the Robinson-Bertotti anti-de Sitter
geometry \cite{rb}
\be
d\bar{s}^2 = -\frac{r^2}{r_0^2} dt^2 + \frac{r_0^2}{r^2} dr^2 + r_0^2
d\Omega^2 \, ,
\ee
where $r_0$ is the extremal radius. After the rescaling (\ref{conformal}) the
two-dimensional metric in null conformal coordinates becomes
\be
\label{v-}
ds^2 = -\frac{2}{l^2q^3} \tilde{x}^2 du dv \, ,
\ee
with $u = v + \frac{l^2q^3}{\tilde{x}}$, $\tilde{x}=l\tilde{\f}$. Proceeding
in a similar way for the near-extremal configuration $v>v_0$, we obtain
\be
\label{v+}
ds^2 = -(\frac{2}{l^2q^3} \tilde{x}^2-l\Delta m) d\bar{u} dv \, ,
\ee
with
\be
\label{u+}
\bar{u} = v + \sqrt{\frac{2lq^3}{\Delta m}} {\mathrm arctanh}
\sqrt{\frac{2}{l^3q^3\Delta m}} \> \tilde{x} \, .
\ee
The coordinates $(v,u)$ and $(v,\bar{u})$ are the radial null coordinates
corresponding to those of RN
$(t+r^{\ast}, t-r^{\ast})$ before and after the shock wave. Imposing the
continuity of (\ref{v-}) and (\ref{v+}) along $v=v_0$ one obtains
\be
u = v_0 + a \> {\mathrm cotanh} \frac{\bar{u}-v_0}{a} \, ,
\ee
where $a = \sqrt{\frac{2lq^3}{\Delta m}}$. From this relation we can work out
immediately the outgoing energy flux of Hawking radiation in terms of the
Schwarzian derivative between the coordinates $u$ and $\bar{u}$
\be
\langle T^f_{\bar{u}\bar{u}} \rangle = -\frac{1}{24\pi} \{ u, \bar{u} \} =
\frac{1}{12\pi a^2} = \frac{\pi}{12} T_H^2 \, .
\ee
We observe that this flux is constant and coincides with the thermal value of
Hawking flux for near-extremal RN black holes, where $T_H$ is Hawking's
temperature. This fact can be understood easily since the AdS$_2\times$S$^2$
geometries associated to (\ref{v-}) and (\ref{v+}) represent indeed the
near-horizon limit of the RN geometries (\ref{rn2})
due to the shock wave (\ref{shock}). So the constant thermal flux for every
value of $\bar{u}$ corresponds to the flux measured by an inertial observer
at future null infinity approaching the event horizon of the RN black hole.
In the light of this remark it is interesting to point out that the JT theory
also describes the dynamics of extremal and near-extremal RN black holes close
to the horizon in the presence of a spherically symmetric Klein-Gordon field.
This is so because, as it is well-known, the scalar field propagates freely
near the horizon.\\

Our purpose now is to analyze the back-reaction effects in the evaporation
process. The one-loop effective theory is obtained by adding the non-local
Liouville-Polyakov term to the classical action. We then get ($\l^2=l^{-2}
q^{-3}$)
\bea
\label{paction}
I &=& \int d^2x \sqrt{-g} \left[ R \tilde{\phi} + 4 
\lambda^2 \tilde{\phi} -\frac{1}{2} \sum_{i=1}^N |\nabla f_i|^2 \right]
\nonumber \\
&-&  \frac{N\hbar}{96\pi} \int d^2x \sqrt{-g} R \; \square^{-1} R + \xi
\frac{N\hbar}{12\pi} \int d^2x \sqrt{-g} \lambda^2 \, ,
\eea
where we have considered the presence of $N$ scalar fields. The parameter $N$
allows us to consider the theory in the large $N$ limit, keeping $N\hbar$
fixed. Moreover we have also added a local conterterm (in the form of a 2d
cosmological constant which corresponds to the freedom of
adding a constant to the 2d conformal anomaly), mimicking the analysis
of dilaton gravity theory \cite{rst}, to ensure that the extremal geometry
remains an exact solution of the one-loop theory at $\xi=1$. Our results are
the same irrespective of the value of $\xi$. We should mention now that for the
region we are interested in, the conformal factor $\sqrt{\f}$ of
(\ref{conformal}) is almost constant and therefore the semiclassical
quantization in terms of the Einstein-Maxwell action and the JT action are
equivalent. Far from the horizon this is no more true.
The equations of motion derived from (\ref{paction}) in conformal gauge
$ds^2=-e^{2\r}dx^+dx^-$ are (from now on we take $\xi=1$)
\bea
\label{eq1}
2\pmas \pmen \r + \l^2 e^{2\r} &=& 0 \, , \\
\label{eq2}
\pmas \pmen \tilde{\f} + \l^2 \tilde{\f} e^{2\r} &=& 0 \, , \\
\label{eq3}
\pmas \pmen f_i &=& 0 \, , \\
\label{eq4}
-2\partial^2_{\pm} \tilde{\f} + 4 \partial_{\pm} \rho \partial_{\pm} 
\tilde{\f} &=& T^f_{\pm \pm} - \frac{N\hbar}{12\pi} t_{\pm} - \\
&& \frac{N\hbar}{12\pi} \left( (\partial_{\pm} \r )^2 -
\partial_{\pm}^2 \r \right) \, . \nonumber
\eea
The functions $t_{\pm}(x^{\pm})$ are related with the boundary conditions of
the theory and depend on the quantum state of the system. The Liouville
equation (\ref{eq1}) has the general solution
\be
\label{lmetric}
ds^2 = -\frac{\pmas A_+ \pmen A_-}{(1+\frac{\lambda^2}{2} A_+ A_-)^2} dx^+
dx^- \, ,
\ee
\begin{figure}
\centerline{\psfig{figure=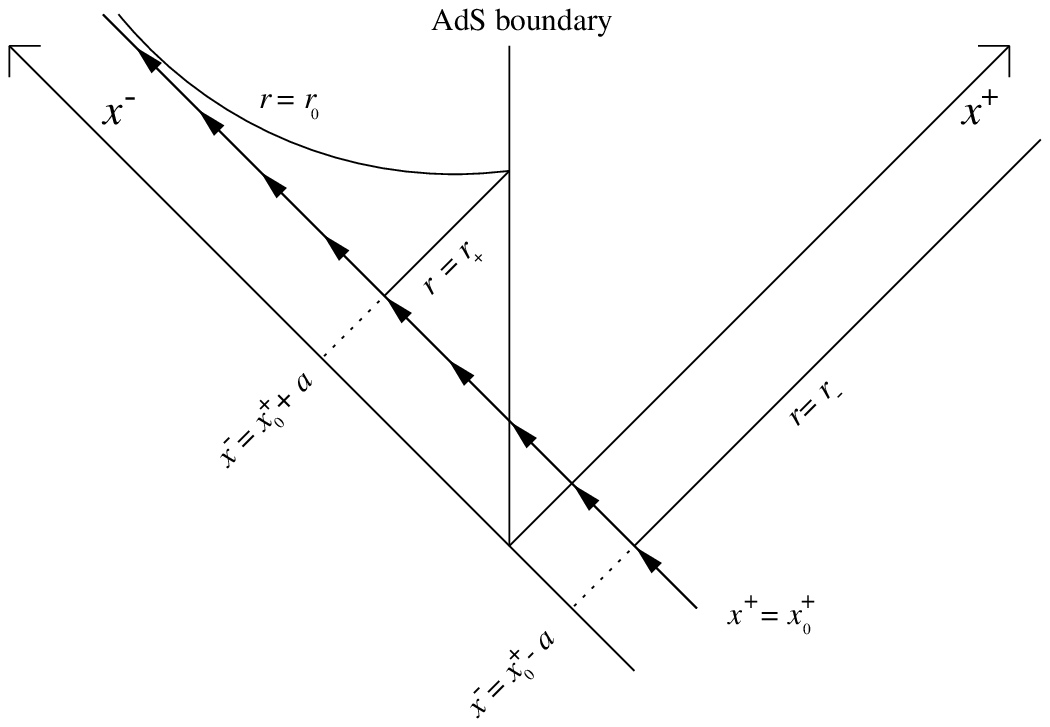,width=3.in,angle=-90}}
\end{figure}
\begin{center}
\makebox[8.5cm]{\parbox{8.5cm}{\small \noindent FIG.1. Kruskal diagram of
near-extremal black hole. The two timelike boundaries of near-horizon geometry
AdS$_2$ are represented by the vertical line $x^-=x^+$. The infalling shock
wave emerges from one boundary (left side of $x^-=x^+$ line) and, crossing the
outer and inner horizons, reaches the other boundary (right side of
$x^-=x^+$ line).}}
\end{center}
where $A_{\pm}(x^{\pm})$ are arbitrary chiral functions. We can choose a
particular form of the functions $A_{\pm}$ as a way to fix completely the
conformal coordinates. We find convenient to choose
\be
\label{gauge}
A_+ = x^+ \, , \qquad
A_- = \frac{-2}{\l^2 x^-} \, .
\ee
Before the shock wave these coordinates $x^{\pm}$ correspond to the RN
coordinates $(v,u)$ and after ($v>v_0$) the relation is
\be
\label{clas1}
v = x_0^+ + a \> {\mathrm arctanh} \frac{x^+-x_0^+}{a} \, , \\
\ee
together with $u=x^- $.
Then both metrics (\ref{v-}) and (\ref{v+}) are brought into (\ref{lmetric})
and the physical information is encoded in the field $\tilde{\f}$. At the
classical level the solution for it is given by
\be
\tilde{\f} = lq^3 \frac{1- \Theta(x^+-x_0^+) \frac{\Delta m}{2lq^3}
(x^+-x_0^+) (x^--x_0^+)}{x^--x^+} \, .
\ee
After the shock wave ($x^+>x_0^+$) the extremal radius is given by curve
$\tilde{\f}=0$: $(x^+-x_0^+)(x^--x_0^+) = a^2$, and the outer and inner
apparent horizons $r=r_{\pm}$ are given by the condition
$\partial_+ \tilde{\f} =0$: $x^- = x_0^+ \pm a$. The corresponding Kruskal
diagram (in this region the coordinates $x^{\pm}$ are regular at the horizon
and therefore they represent a sort of Kruskal frame) is given by Fig.1.\\

At the quantum level we have to solve equations (\ref{eq1}-\ref{eq4}) and the
crucial point is to choose the adequate functions $t_{\pm}(x^{\pm})$ for the
physical situation. The natural choice is $t_v=t_{x^+}=0$ and $t_u= t_{x^-}=0$
before the shock-wave and $t_{x^+}=\frac{1}{2} \{ v, x^+ \}$ and $t_{x^-}=0$
after, where now $v$ is the light-cone coordinate of the evaporating
Vaidya-type metric
\be
\label{v++}
ds^2 = -(\frac{2\tilde{x}^2}{l^2q^3}-lm(v)) dv^2 + 2d\tilde{x} dv \, .
\ee
The remarkable property of the equations of the near-horizon effective theory
is that one can solve them also in conformal gauge. We find that
\be
\tilde{\phi} = \frac{F(x^+)}{x^--x^+} + \frac{1}{2}  F'(x^+) \, ,
\ee
$\tilde{x} = l \tilde{\f}$, where the function $F(x^+)$ satisfies the
following differential equation
\be
 F'''= \frac{N\hbar}{24\pi} \left( -\frac{F''}{F} +
\frac{1}{2} \left( \frac{F'}{F} \right)^2 \right) \, ,
\ee
and relates the $x^+$ and $v$ coordinates
\be
\frac{dv}{dx^+} = \frac{lq^3}{F} \, .
\ee
The evaporating mass is then given by
\be
m(x^+) = \frac{24\pi }{ N\hbar lq^3} F^2  F''' \, ,
\ee
and can be related to the boundary function $t_{x^+}$
\be \label{tp}
t_{x^+} = \frac{lq^3m(x^+)\Theta(x^+-x_0^+)}{2F^2} \, .
\ee
The fact that the functions $t_{\pm}$ can be discontinuous for coordinates
$x^{\pm}$ associated to free (or Liouville) fields was pointed out in
\cite{cfn}. The expression (\ref{tp}), and also the function $F(x^+)$,  admits a 
series expansion in powers
of $\hbar$, where the classical term, obtained using the classical relation
(\ref{clas1}), is given by 
\be
\label{boco}
t_{x^+} = \frac{a^2 \Theta(x^+-x_0^+)}{(a^2-(x^+-x_0^+)^2)^2} \, \ .
\ee
As before, the curve $\tilde{\f}=0$
\be \label{aor}
x^-=x^+ - \frac{2F}{F'} \, ,
\ee
represents the location of the extremal
radius and $\partial_+ \tilde{\f} =0$ defines the inner and outer apparent 
horizons in the spacetime of the evaporating black hole
\be \label{por}
x^-=x^+ -\frac{F'}{F''} \pm\frac{\sqrt{F'^2 -2F F''}}{F''} \, .
\ee 
The intersection of these three curves takes place when 
\be
\label{ep}
F'^2 - 2FF''=2lq^3m(x^+)=0 \, ,
\ee
i.e. at the end of the evaporation. 
On the other hand , it is easy to show that 
$\pmas m(x^+)=-\frac{N\hbar}{24\pi}\frac{m(x^+)}{F}$ which can be readily
integrated in $v$ coordinate giving $m(v)=\Delta m e^{-\frac{N\hbar}{24\pi
lq^3}(v-v_0)}$ and so $m=0$ is given by $v=+\infty$ and not before (had we
started with the classical boundary term (\ref{boco}) we would have obtained
a finite evaporation time). From the numerical graph of the function $F(x^+)$
it is clear that this happens at a finite 
\begin{figure}
\centerline{\psfig{figure=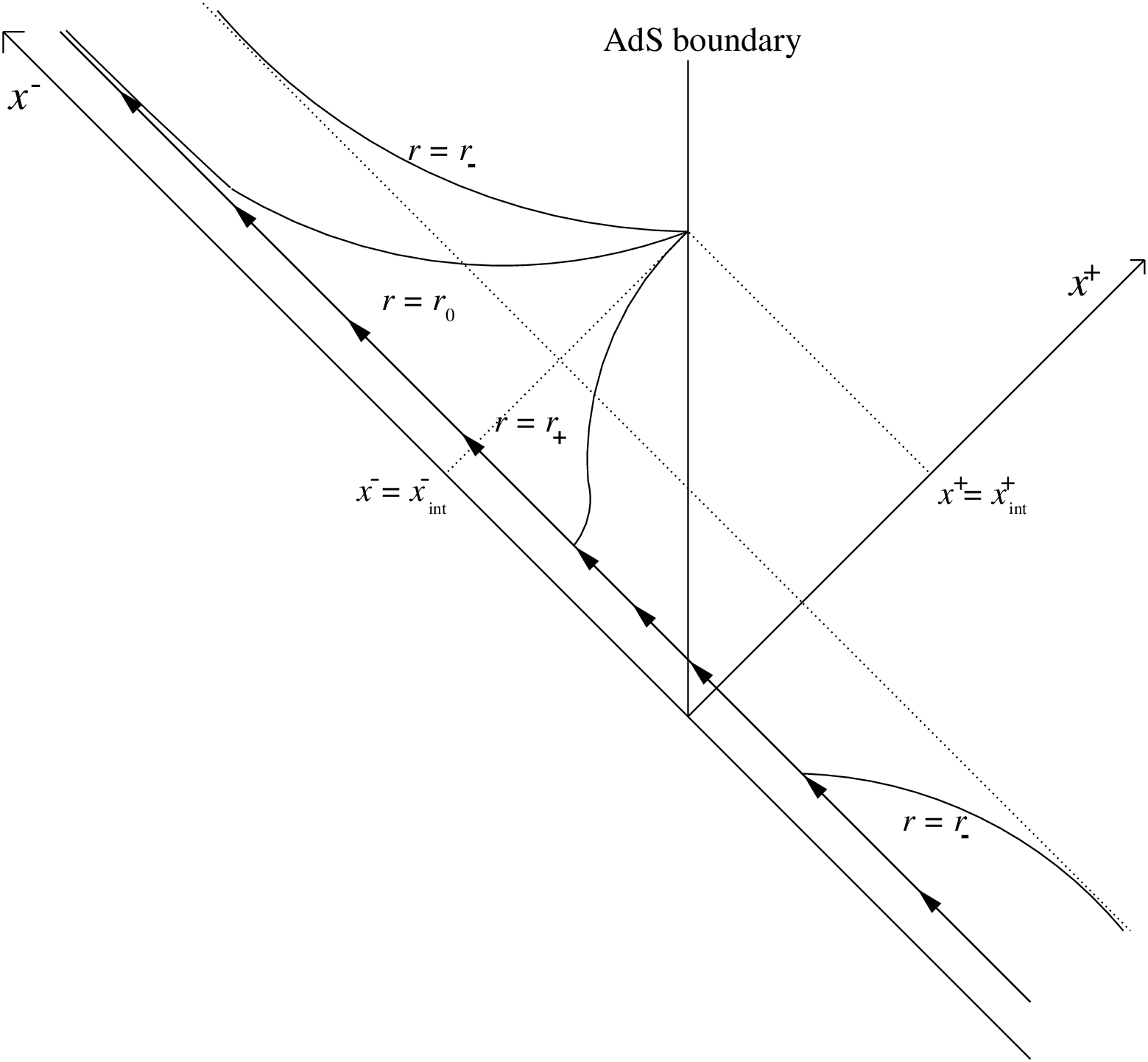,width=3.in,angle=-90}}
\end{figure}
\begin{center}
\makebox[8.5cm]{\parbox{8.5cm}
{\small FIG.2. Kruskal diagram of semiclassical evolution of
RN black hole. The outer apparent horizon shrinks until it meets both inner
horizon and extremal radius curve at the AdS boundary.}}
\end{center}
value of $x^+=x^+_{int}$ for which $F(x^+_{int})=0$ and moreover
from eq. (\ref{ep}) it also follows that $F'(x^+_{int})=0$. One can also show
that $F''(x^+_{int})$ is nonzero while all the other derivatives
vanish, so that locally close to the intersection
point $F(x^+)$ behaves as a parabola with exponentially suppressed corrections.
This is enough to prove , see (\ref{aor}) and (\ref{por}), that the
intersection point belongs to the AdS boundary, i.e. $x^+_{int}=x^-_{int}$. 
The graph of all these curves is shown in Fig. 2, where the saddle point in 
the outer apparent horizon curve $r_+$ signals the transition 
from the strong to the weak back-reaction regimes as described in \cite{lo}. 
At the end-point the curves $\f=0$ and $\partial_+\f=0$ are null and the
dilaton function is well represented asymptotically by the extremal solution
\be 
\f = \frac{\pmas^2
F(x^+_{int})}{2}\frac{(x^+-x_{{\mathrm int}}^+)(x^--x_{{\mathrm int}}^+)}
{x^--x^+} + ....
\ee
where the dots are nothing but exponentially small corrections.  
So, as time passes the solution becomes 'more and more extremal' but without 
actually coming back to the extremal state due to the infinite evaporation
time. Due to the fact that $T_H=0$ for the extremal black hole our
exact results are in  agreement with the third law of thermodynamics applied to
black holes (see for instance \cite{isr}), despite the fact that the weak
energy condition is violated close to the horizon, as well as with those
obtained using the adiabatic approximation \cite{st}.\\

This research has been partially supported by the CICYT and DGICYT, Spain.
D. J. Navarro acknowledges the Ministerio de Educaci\'on y Cultura for a FPI
fellowship. A.F. thanks R. Balbinot for useful discussions and the Department
of Theoretical Physics of Valencia University for hospitality during the late
stages of this work. D.J.N. and J.N-S. also wish to thank J. Cruz and P. 
Navarro for comments.

\vspace{0.5cm}
\noindent $^{\ast}$Email address: fabbria@bo.infn.it\\
\noindent $^{\dagger}$Email address: dnavarro@ific.uv.es\\
\noindent $^{\ddagger}$Email address: jnavarro@lie.uv.es 
%%%%%%%%%%%%%%%%%%%%%%%%%%%%%%%%%%%%%%%%%%%%%%%%%%%%%%%%%%%%%%%%%%%%%%%%%%%%

\end{multicols}

\vspace{-2.2cm}
\begin{thebibliography}{99}

\bibitem{h}
S.W. Hawking, {\it Nature} {\bf 248}, 30 (1974); {\it Comm. Math. Phys.}
{\bf 43}, 199 (1975).

\bibitem{svms}
A. Strominger and C. Vafa, {\it Phys. Lett.} {\bf B379}, 99 (1996);
C.G. Callan and J.M. Maldacena, {\it Nucl. Phys.} {\bf B472}, 591 (1996);
J. Maldacena and A. Strominger, {\it Phys. Rev. Lett.} {\bf 77}, 428 (1996).

\bibitem{ghs}
D. Garfinkle, G.T. Horowitz and A. Strominger, {\it Phys. Rev.} {\bf D43},
3140 (1991).

\bibitem{cghs}
C.G. Callan, S.B. Giddings, J.A. Harvey and A. Strominger, {\it Phys. Rev.}
{\bf D45}, R1005 (1992).

\bibitem{rst}
J.G. Russo, L. Susskind and L. Thorlacius, {\it Phys. Rev.} {\bf D46}, 3444
(1993); {\it Phys. Rev.} {\bf D47}, 533 (1993).

\bibitem{h2}
S.W. Hawking, {\it Phys. Rev. Lett.} {\bf 69}, 406 (1992).

\bibitem{bpp}
S. Bose, L. Parker and Y. Peleg, {\it Phys. Rev. Lett.} {\bf 76}, 861 (1996).

\bibitem{mik}
A. Mikovic, {\it Class. Quant. Grav.} {\bf 13}, 209 (1996).

\bibitem{tgh}
L. Thorlacius, {\it Nucl. Phys. Proc. Suppl.} {\bf 41}, 245 (1995);
S.G. Giddings, {\it ``Quantum mechanics of black holes''}, hep-th/9412138;
A. Strominger, {\it ``Les Houches lectures on black holes''}, hep-th/9501071.

\bibitem{st}
A. Strominger and S.P. Trivedi, {\it Phys. Rev.} {\bf D48}, 5778 (1993).

\bibitem{lo}
D.A. Lowe and M. O'Loughlin, {\it Phys. Rev.} {\bf D48}, 3735 (1993).

\bibitem{jt}
R. Jackiw, in {``Quantum Theory of Gravity''}, edited by S.M. Christensen
(Hilger, Bristol, 1984), p.~403; C. Teitelboim, in {\it op. cit.}, p.~327.

\bibitem{mw}
J.M. Maldacena, {\it Adv. Theor. Math. Phys.} {\bf 2}, 231 (1998); E. Witten,
{\it Adv. Theor. Math. Phys.} {\bf 2}, 253 (1998).

\bibitem{nnn}
J. Navarro-Salas and P. Navarro, hep-th/9910076 (to appear in {\it Nucl. 
Phys.} {\bf B}); D.J. Navarro, J. Navarro-Salas and P. Navarro, hep-th/9911091
(to appear in {\it Nucl. Phys.} {\bf B}).

\bibitem{rb}
I. Bobinson, {\it Bull. Akad. Pol.} {\bf 7}, 351 (1959); B. Bertotti, {\it
Phys. Rev.} {\bf 116}, 1331 (1959).

\bibitem{cfn}
J. Cruz, A. Fabbri and J. Navarro-Salas, {\it Phys. Lett.} {\bf B449}, 30
(1999).

\bibitem{isr}
W. Israel, {\it Phys. Rev. Lett.} {\bf 57}, 397 (1986).

\end{thebibliography}
\end{document}